\documentclass[twocolumn,twoside,slac]{revtex4}
\usepackage{graphicx}
\usepackage{fancyhdr}
\renewcommand{\headrulewidth}{0pt}
\renewcommand{\footrulewidth}{0pt}

\begin{document}

\title{FAYE: A Java Implement of the Frame/Stream/Stop Analysis Model.}

%

\author{S. Patton}
\affiliation{LBNL, Berkeley, CA 94720, USA}

\begin{abstract}
FAYE, The Frame AnalYsis Executable, is a Java based implementation of
the Frame/Stream/Stop model for analyzing data. Unlike traditional Event
based analysis models, the Frame/Stream/Stop model has no preference as
to which part of any data is to be analyzed, and an Event get as equal
treatment as a change in the high voltage. This model means that FAYE is
a suitable analysis framework for many different type of data analysis,
such as detector trends or as a visualization core. During the design of
FAYE the emphasis has been on clearly delineating each of the
executable's responsibilities and on keeping their implementations as
completely independent as possible. This leads to the large part of FAYE
being a generic core which is experiment independent, and smaller
section that customizes this core to an experiments own data structures.
This customization can even be done in C++, using JNI, while the
executable's control remains in Java. This paper reviews the
Frame/Stream/Stop model and then looks at how FAYE has approached its
implementation, with an emphasis on which responsibilities are handled
by the generic core, and which parts an experiment must provide as part
of the customization portion of the executable.
\end{abstract}

\maketitle

\thispagestyle{fancy}


\section{Introduction}

The Frame/Stream/Stop model developed for CLEO~III analysis was
originally developed in C++~\cite{chep97-paper}. Since then is has been
successfully deployed in a number of rolls at CLEO. Meanwhile the
IceCube experiment, which has adopted Java as its main language for DAQ
and Data Handling, wanted to use the Frame/Stream/Stop model in its
code. This meant that a Java version of the model needed to be
developed.

\section{The Frame/Stream/Stop Model}

Before examining the Java implementation of the Frame/Stream/Stop model
it is worthwhile reviewing the model itself. The core idea of the model
is that any analysis of data taken from a H.E.P. experiment is
essentially based upon an ``electronic picture'' of the experiment at
certain moments in time. This ``picture'' is made up of various
different elements which change over time and, most importantly, change
at different rates, e.g. a detector's geometry should change much less
frequently than its high voltage (HV) status. Given these ideas the
following components are defined in the Frame/Stream/Stop model.

\begin{description}
\item[Record] This is a set of related data, all of which all will
conceptually change at the same time.
\item[Frame] The ``electronic picture'' of the experiment that is
composed of different types of Records, all of whom are related to the
same time.
\item[Stream] A set of Record, all of the same type, from different
times.
\item[Stop] The occurrence of a new Record in a Stream ``of interest''. When this
occurs during execution the Frame corresponding to the Stop is passed to
analysis routines for processing.
\item[Active Stop] A Stop which occurs on a sequential Stream (the Stream
does not have to be ordered).
\item[Passive Stop] A Stop which occurs in response to, and precedes, an
Active Stop. i.e. if the Record in some Stream of interest changes when
an Active Stop occurs, the The Frame corresponding to this Stream of
interest is passed for processing before the Frame corresponding to the
Active Stop is supplied.
\end{description}

\begin{figure*}
\centering
\includegraphics{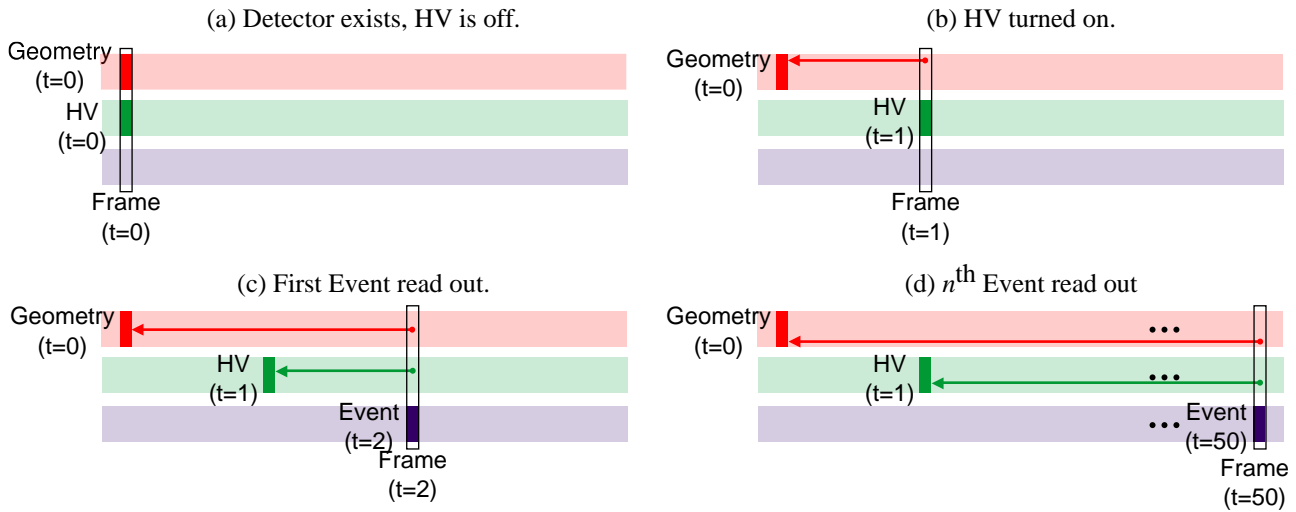}
\caption{The ideas of a Record, a Stream and a set of Frames.}
\label{building-frames}
\end{figure*}

Figure~\ref{building-frames} shows off the ideas of a Record, a Stream
and a set of Frames. In this Figure the three horizontal bands, from top
to bottom, in each diagram represent the Geometry, HV and Event Streams.
The solid blocks of color in these Stream represent an appropriate
Record in that Stream, with the label indicating the time associated
with that Record. The tall black open rectangle represents a Frame which
contained either the Records corresponding to the time of the Frame, the
most recent Record in a Stream which has no Record matching the time of
the Frame (as seen for the Geometry Stream in
Figure~\ref{building-frames}(b)), or no Record for a Stream that has no
record prior to the time of the Frame (as seen for the Event Stream in
Figure~\ref{building-frames}(b)).

\begin{figure*}
\centering
\includegraphics{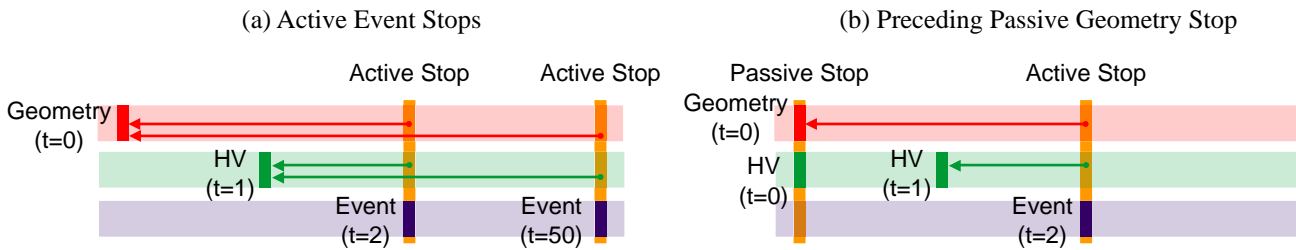}
\caption{An illustration of the difference between Active and Passive
Stops.}
\label{active-passive}
\end{figure*}

Figure~\ref{active-passive} helps demonstrate the difference between
Active and Passive Stops using the example of an Event Display. In this
case the Event and Geometry Streams are the Streams of interest for this
executable. The executable is driven by a sequence of Event Records each
of which generate an Active Stop. The Geometry Stream Records, on the
other hand, are not provided sequentially but rather are read from a
database to fill in the Frames needed for the Active Stops. As can be
seen in the Figure, before the first Active Stop can be supplied, a
Passive Stop cause by the reading of the initial Geometry Record should
be supplied to the analyses so that they can process this initial
geometry before processing the first Event.

\section{Implemention of the Frame/Stream/Stop model}

The Java implementation of the Frame/Stream/Stop model is broken down
into three separate layers. The first and most general layer is a
generic record processing loop. This part of the implementation is based
on the classic source/listener pattern that is used throughout the
standard Java libraries. In fact, this layer is so general that it is
scheduled to become part of the {\tt freehep} Java
library~\cite{freehep-site}

The next layer implements the major ideas of the Frame/Stream/Stop model
and, as such, contains the logic needed to supply Frames to analyses.
However as the exact definitions of Records and Streams are experiment
dependent these detail are included in the third layer, rather than the
second.

The third layer is the only layer that an experiment needs to provide
to tailor FAYE to work with their experiment. This layer includes the
definition of Records and Streams and also includes the mechanism to
dispatch Frames to the correct analysis routines.

\subsection{The {\tt freehep} Layer}

The generic record loop code is contained in the {\tt
org.freehep.record} packages. The core interface in these part of the
code is the {\tt RecordListener} interface.

\subsubsection{The {\tt RecordListener} interface}

\begin{figure*}
\centering
\includegraphics{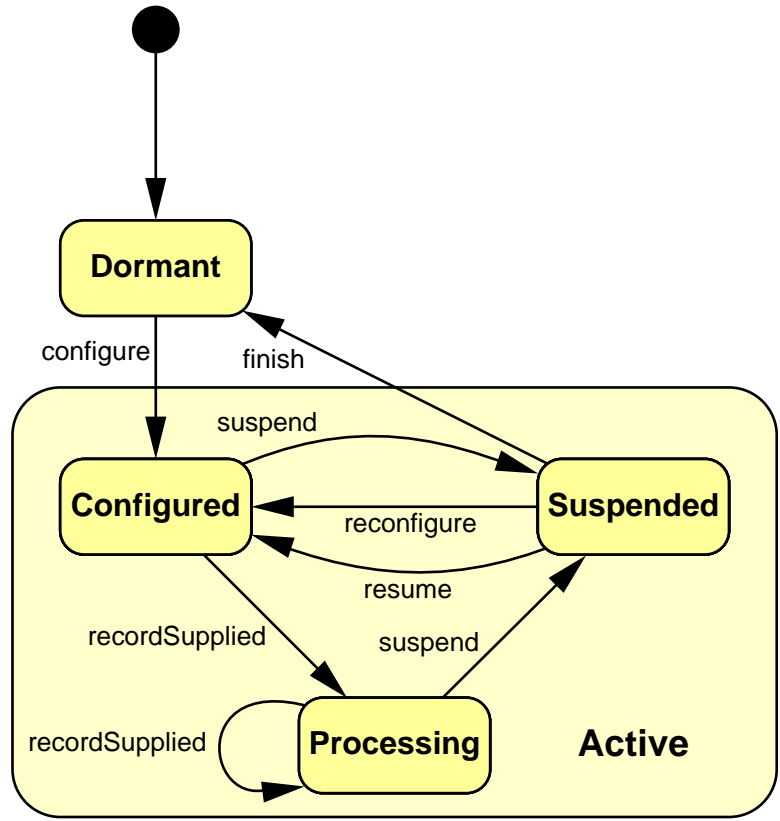}
\caption{The life-cycle of the {\tt RecordListener} interface.}
\label{rl-lifecycle}
\end{figure*}

The {\tt RecordListener} interface defines the methods that any analysis
class must implement in order to be executed as part of a record loop.
Figure~\ref{rl-lifecycle} shows the life-cycle that all implementations
of the {\tt RecordListener} interface are expected to follow.

When a {\tt RecordListener} instance is created it starts off in a {\sf
dormant} state. When a record loop is about to begin, the instance will
receive a {\tt configure} message and transition into the {\sf
configured} state. This transition gives the instance an opportunity to
read and respond to any input parameters that may have been set, and
thus prepare itself for {\tt recordSupplied} messages.

In the course or normal processing a {\tt RecordListener} interface can
expect to receive one or more {\tt recordSupplied} messages after it has
been configured. The first of these messages will cause the instance to
transition into the {\sf processing} state, where it will remain while
it receives more {\tt recordSupplied} events.

Eventually there will be no more records to supply, either because a
user defined limit has been reached or no more Records are available.
When this happens the instance will receive a {\tt suspend} message
which will cause it to transition into the {\tt suspended} state. This
transition allows the instance to release and time-critical resources,
e.g. database locks, so that other jobs can continue while this job is
not processing records.

There are three possible transitions out of the {\sf suspended} state,
two of these transitions return the instance to a {\sf configured}
state, while one returns it to the {\sf dormant} state. The difference
between the {\tt reconfigured} and {\tt resume} messages, both of which
cause a transition into the {\sf configured} state, is that the first of
these implies that the input parameters to the instance of {\tt
RecordListener} may have changed and that the instance should be read
and respond to these new values, while the other message guarantees that
none of these parameters have changed and so the instance can resume
where it left off.

The other transition out of the {\tt suspended} state is caused by a
{\tt finish} message being received. This signals that the instance is
not longer in the set of classes that will be executed if or when
another record loop is executed. This transition allows the instance to
clean up any intermediate data it contains and, if necessary, output a
summary of the processing it has done.

The instance of a {\tt RecordListener} can expect to only be destroyed
from the {\sf dormant} state. This gives the instance one last change to
cleanly shut down. However, as this transition is cause by the {\tt
finalize} Java method it should normally not be used as there is no
determinist way of knowing when or if this method will be called.

The state machine shown in Figure~\ref{rl-lifecycle} leads to the {\tt
RecordListener} interface as declared in Figure~\ref{rl-interface}.

\begin{figure*}
\centering
\begin{verbatim}
public interface RecordListener
        extends EventListener
{
    public void configure(ConfigurationEvent event);
    public void finish(RecordEvent event);
    public void recordSupplied(RecordSuppliedEvent event);
    public void reconfigure(ConfigurationEvent event);
    public void resume(RecordEvent event);
    public void suspend(RecordEvent event);
}
\end{verbatim}
\caption{The {\tt RecordListener} interface.}
\label{rl-interface}
\end{figure*}

\subsubsection{The other classes and interfaces}

The other classes and interfaces in the {\tt org.freehep.record}
packages provide tools that can be used in conjunction with {\tt
RecordListener} interface. The following list details some of the issues
these tools tackle:

\begin{description}
\item[Sequencing] This allows a set of {\tt RecordListener}s to be executed
sequentially.
\item[Branching] This enables two sequences of {\tt RecordListener}s to
be executed independently of each other.
\item[Conditional execution] This allows for preemptive termination of a
sequence of {\tt RecordListener}s when it has been decided that
continued execution would be a waste of time.
\item[Source interfaces] This defined the interfaces which should be
implemented for sequential and interactive processing of the record
loop.
\end{description}

\subsection{The FAYE Layer}

The FAYE layer implements the mechanics needed to run the
Frame/Stream/Stop model on to of the generic record loop. Its
responsibilities lie mainly in the area of the creation of Frames, which
serve as the records in this set-up. This should not be confused with
the idea of Records that are elements of the Frame itself.

This layer provides a {\tt FayeSource} class which is an implementation
of the record source interface that is declared in the {\tt freehep}
layer. This implementation contains a set of {\tt FayeStopSource}
objects, each of which is able to read Record data from a single source,
e.g. a file, database, etc., and provide it to the {\tt FayeSource}
instance so that it can work out which is the next Stop that should be
used to create the next Frame supplied to the analysis. The Frame itself
is created using the {\tt FrameFactory} interface of declared in this
layer. By using a factory interface and leaving the exact implementation
to be contained in the Experiment's layer, an experiment is free to
choose how it wants to access data in the Frame. The created Frame is
returned for use in the generic loop.

The {\tt FayeStopSource} objects contained in the {\tt FayeSource} are
not only responsible for reading Record data from their own sources, but
they are also required to supply to the {\tt FayeSource} object the next
Active Stop they can read and, given an Active Stop, the ``earliest''
Passive Stop they can read. The {\tt FayeStopSource} can also act as a
{\tt RecordListener} so that it can can be managed by the {\tt
RecordListener}'s life-cycle and, more importantly, its {\tt
recordSupplied} implementation can be used to load the Frame with the
appropriate data from its source.

The single {\tt RecordListener} implementation provided by this layer is
the {\tt FayeListener} class. This acts as a two phase listener. During
the first phase, remembering that the new Frame has already been
created, all the {\tt FayeStopSource} objects get an opportunity to add
their data to the Frame. The second phase then supplies this filled
Frame to the analysis {\tt RecordListener} implementations.

\subsection{The Experiment (IceCube) Layer}

\begin{figure*}
\centering
\includegraphics{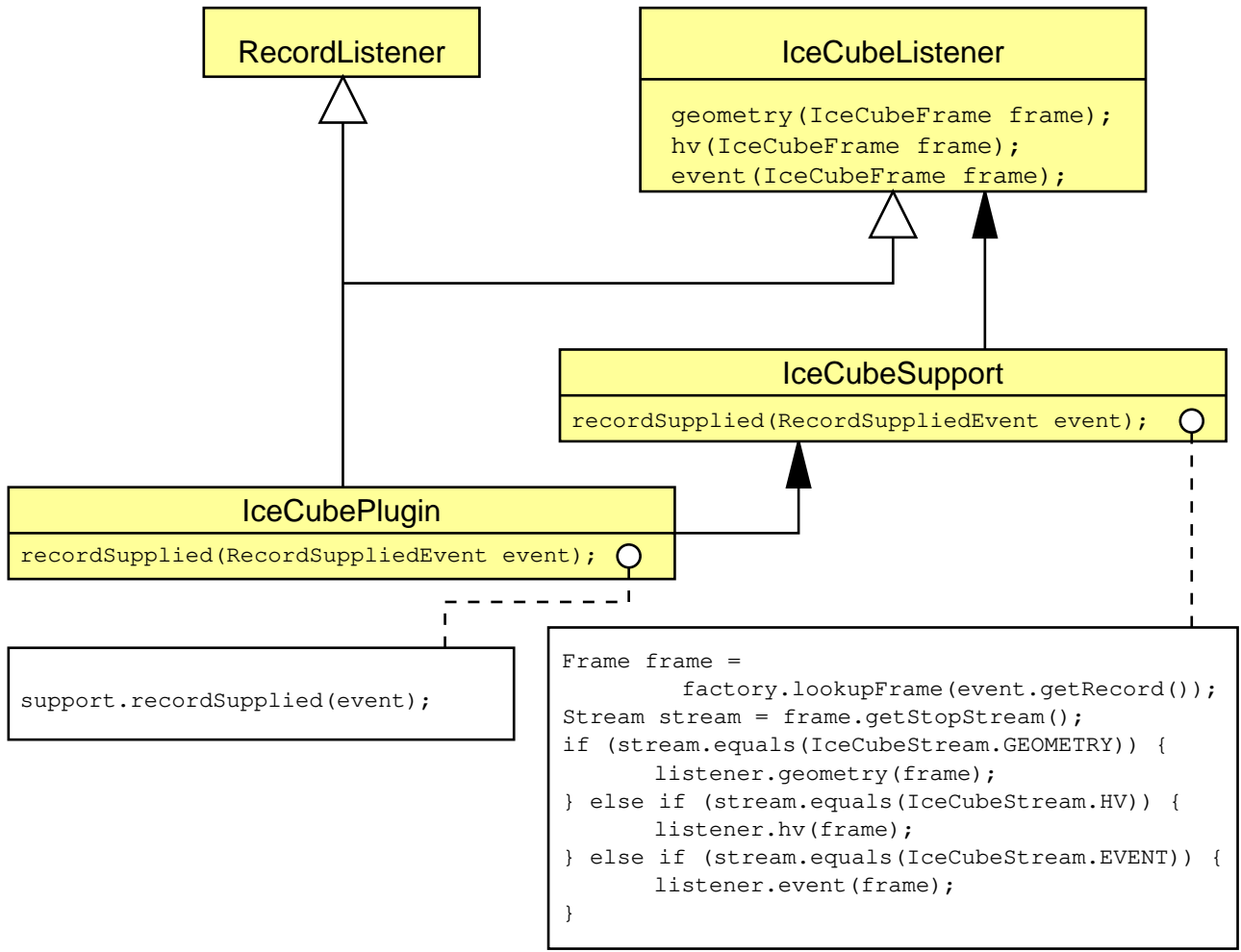}
\caption{The organization of classes in the Experiment's layer.}
\label{icecube-layer}
\end{figure*}

The Experiment's layer allows an experiment to specialize this framework
to it own situation. In this paper we will use the IceCube experiment
as the example experiment. The main aim of the specialization is to
enable the inclusion of the experiments own Streams into the frameworks
and handle the dispatch of the Frames to the correct methods.
Figure~\ref{icecube-layer} shows the standard way of implementing this.
The {\tt IceCubeListener} class defines methods for each of the standard
Streams in the experiment. The {\tt IceCubePlugin} class bring together
the generic {\tt RecordListener} interface and experiment specific one
and handles the dispatch of the Frames to the right method by
using an instance of the {\tt IceCubeSupport} class. This class, in
turn, implements its {\tt recordSupplied} method as a large switch
statement that matches the Stream which caused the Frame to be supplied
to the matching method in the {\tt IceCubeListener} interface.

\section{Summary}

The Frame/Stream/Stop model provides a flexible framework in which to
develop H.E.P. analyses. This has been demonstrated by its C++
implementation at CLEO.

The Java implementation of this model is based on a {\tt freehep}
foundation, which means that it can easily be used elsewhere, for
example as part of JAS3~\cite{jas-site}.

An experiment only needs to specialize around half a dozen classes to
tailor this framework to it own situation.

\end{document}